\documentclass[a4paper,onecolumn,11pt,accepted=2022-05-17]{quantumarticle}
\pdfoutput=1
\usepackage[utf8]{inputenc}
\usepackage[english]{babel}
\usepackage[T1]{fontenc}
\usepackage{amssymb, amsmath}
\usepackage{hyperref}

\usepackage{tikz}
\usepackage{lipsum}

\begin{document}

\title{Solving the Bose-Hubbard model in new ways}
\author{Artur Sowa}
\email{sowa@math.usask.ca}
\affiliation{Department of Mathematics and Statistics, University of Saskatchewan, Canada}
\author{Jonas Fransson}
\affiliation{Department of Physics and Astronomy, University of Uppsala, Sweden}
\maketitle

\begin{abstract}
We introduce a new method for analysing the Bose-Hubbard model for an array of boson sites with nearest neighbor interactions. It is based on a number-theoretic implementation of the  creation and annihilation operators that constitute the model. One of the advantages of this approach is that it facilitates accurate computations involving multi-particle states.  In particular, we provide a rigorous computer assisted proof of quantum phase transitions in finite systems of this type. 

Furthermore, we investigate properties of the infinite array via harmonic analysis on the multiplicative group of positive rationals. This furnishes an isomorphism that recasts the underlying Fock space as an infinite tensor product of Hecke spaces, i.e., spaces of square-integrable periodic functions that are a superposition of non-negative frequency harmonics. Under this isomorphism, the number-theoretic creation and annihilation operators are mapped into the Kastrup model of the harmonic oscillator on the circle. It also enables us to highlight a kinship of the model at hand with an array of spin moments with a local anisotropy field. This identifies an interesting physical system that can be mapped into the model at hand.
\end{abstract}

  \vspace{.2cm}


\section{Introduction}

The Bose-Hubbard Hamiltonian assumes the form
\begin{equation}\label{BH}
\mathcal{H}= \sum_{n=1}^{\infty}  \frac{U}{2} \, \hat{N}_n(\hat{N}_n - 1) - \mu\, \hat{N}_n
-t\, (\hat{a}_{n}^\dagger \hat{a}_{n+1} + \hat{a}_{n+1}^\dagger \hat{a}_n ),
\end{equation}
where the number operators are defined via $\hat{N}_n = \hat{a}_{n}^\dagger \hat{a}_{n}$.  The first two terms account for the sites' Coulomb potential $U$ (attractive if $U>0$ and repulsive otherwise). The constant $\mu$ is the chemical potential. The last term  is responsible for disorder. The hopping amplitude $t$ controls the relative energy of disorder. Perhaps the fact that summation is over $n\in \mathbb{N}$ rather than $n\in \mathbb{Z}$ is not exactly standard. This constraint may be interpreted as the existence of an edge.

In preexisting physics literature, the method constituting state-of-the-art when it comes to studies of the Bose-Hubbard model is density matrix renormalization group (DMRG), see for instance \cite{PhysRevLett.69.2863,PhysRevB.58.R14741,PhysRevB.61.12474,PhysRevLett.95.240404,PhysRevB.101.075419}. While density matrix renormalization group is considered to be the method of reference, methods like dynamical mean field theory (DMFT) \cite{PhysRevB.80.245110} and quantum Monte-Carlo (QMC) \cite{PhysRevB.100.174517} are also currently used to provide a complementary view of the physics contained in the Bose-Hubbard model. Specifically, using the density matrix renormalization group theory, it is possible to go beyond the basic set-up of the Bose-Hubbard model, Eq. (\ref{BH}), with additions of inter-site Coulomb interactions, e.g., \cite{PhysRevB.61.12474}, and spin interactions, e.g., \cite{PhysRevLett.95.240404}, which makes the modelling more versatile in its applicability and also enriches the phase space of fundamentally interesting physics captured through the model. Shortcomings with density matrix renormalization group are, however, that the method is all numerical based and that it reduces the intrinsic phase space by truncating high energy modes out of the computations. While the first issue can be overcome by continuously increasing computing capabilities which allows both computational refinement as well as increasing system size, the second issue is of a more fundamentally problematic nature. As the method is based on the idea of repeated truncations of the high energy degrees of freedom to the benefit of capturing the low energy physics, there is certainly an inherent risk of losing phases in the phase space which can only be obtained when considering the full picture of the model.
It is  noteworthy that this has been accomplished for the particular case of the two-site Bose Hubbard model. It has been solved exactly \cite{Links} and shown to exhibit rich quantum dynamics \cite{Tonel}. The experimental relevance of the two-site model has been discussed in-depth in \cite{Batchelor}. Other types of two-site models have received close attention as the theoretical foundation for experimental implementation of mesoscopic Schr\"{o}dinger cat states, \cite{Cirak}, and appeared in studies of the quantum phase transitions, \cite{Pan}. There are notable contributions focusing about the nonlinear semi-classical approximations of the Bose-Hubbard model and related fundamental questions, e.g., \cite{Oelkers}, \cite{Javanainen}, \cite{Buonsante}.

Here, we address the Bose-Hubbard model by means of a number theoretic framework which provides access to the exact numerical solutions of the model. Without changing the structure of the Hamiltonian (\ref{BH}), we will examine its number-theoretic implementation, namely:
\begin{equation}\label{NT_BH}
  \mathcal{H}= \sum_{n=1}^{\infty}  \frac{U}{2}\, \hat{N}_{p_n}(\hat{N}_{p_n} - 1) - \mu\, \hat{N}_{p_n} - t\, (\hat{a}_{p_n}^\dagger \hat{a}_{p_{n+1}} + \hat{a}_{p_{n+1}}^\dagger \hat{a}_{p_n}),
\end{equation}
where the $p_1, p_2, p_3,\ldots $ are the consecutive primes. 
The question is how to define $\hat{a}_{p_{n}}$ (a fortiori $ \hat{a}_{p_n}^\dagger$) as an operator on the bosonic Fock space, and ensure that the bosonic Canonical Commutation Relations (BCCR) are satisfied, i.e.
\begin{equation}\label{bCCR}
[\,  \hat{a}_p, \hat{a}_q^\dagger \, ] = \delta_{p,q}, \quad  [\,  \hat{a}_p, \hat{a}_q \, ] = 0\quad \mbox{ for all primes } p,q.
\end{equation}
The answer utilizes observations in the seminal article of J.-B. Bost and A. Connes, \cite{Bost_Connes} on phase transitions in bosonic quantum field theory. The main fact is that the space of square-integrable arithmetic functions, $\ell_2(\mathbb{N})$, is equivalent with the bosonic Fock space. This enables suitable definitions of the creation and annihilation operators, (\ref{on_arithmetic}). We emphasize that they are equivalent with the standard ones, i.e. those expressed in the canonical Fock space formalism, e.g. \cite{Derezinski}, cf. \emph{Example} in the closing of Subsection \ref{subsec_a_ahat}. Crucially, however, when $\ell_2(\mathbb{N})$ is adopted as the model for the Fock space, the Hamiltonian (\ref{NT_BH}) is explicitly and naturally organized as an infinite matrix. Moreover, the entries of the Hamiltonian matrix are computable with perfect accuracy\footnote{Computation of the entries requires integer factorization, which is believed to be NP-hard. However, in practice, factorization of numbers surpassing the Avogadro number can be accomplished in relatively short time in standard platforms, such as Matlab. A more restrictive bottleneck of the  computational approach is determining the eigenvalues and eigenvectors of large matrices which, of course, is a universal problem of numerical analysis.}. This has two main consequences that we explore in this paper: First, one can pass on to exactly computable finite-dimensional models retaining multiparticle phenomena within the scope. Second, the infinite-dimensional Hamiltonian set in this way is amenable to transforming via the generalized Fourier transform for the multiplicative group of positive rationals. This reveals equivalence of the number-theoretic creation and annihilation operators with those inherent in the Kastrup model of the harmonic oscillator on the circle, \cite{Kastrup}. That in turn enables observation of the equivalence of the Bose-Hubbard array with an anisotropic spin chain.

The characteristic features of the number-theoretic implementation are discussed in detail in Section \ref{Section_NT}. Quite crucially for our purposes, $\ell_2(\mathbb{N})$ admits a special orthonormal basis that consists of all point-measures $\delta_n$, $n \in \mathbb{N}$. In this representation the single-particle wave-functions correspond to $\delta_p$ for $p \in \mathcal{P}$ (the set of all primes). This setting makes it easier to discern (theoretically and computationally)  multi-particle states. This has lead us to new insights into the physics of the Bose-Hubbard system, described in Section \ref{Section_numerical}. Specifically, we find that the system admits phases with the distinguished complexity of the ground states, e.g. for some regions in the $\mu, t$ plane (with $U$ fixed) the ground state will be a superposition of single-particle states. In some other regions the ground state is a superposition of multi-particle states. Remarkably, these regions are relatively ample, possessing nonempty interiors, so that quality of the ground state is typically stable under small perturbations of parameters. Furthermore,  computation of the singularities of the \emph{grand canonical ensemble partition function} furnishes rigorous identification of quantum phase transitions in the model. We stress that our method renders the exact Hamiltonian matrix. Hence, the problem of solving the model is reduced to an eigenvalue problem which can be solved numerically with arbitrary accuracy, only limited by the computing capacity (errors are typically less than $10^{-16}$).

In Section \ref{Section_FT_n_Kast} we discuss theoretical aspects of the infinite array of sites. Our main tool is harmonic analysis on the multiplicative group of positive rationals, denoted $\mathbb{Q}_+$. The Pontryagin dual of this group is the infinite-dimensional torus $\hat{\mathbb{Q}}_+ = \prod_{p\in \mathcal{P}} U(1)$, where $\mathcal{P}$ is the set of primes. In this way, we demonstrate the equivalence of the number-theoretic creation/annihilation operators with those constructed by H.A. Kastrup in \cite{Kastrup}, which he did to model the harmonic oscillator in $L_2(U(1))$. As an application, in Section \ref{sec-physical} we uncover certain analogies of our model, i.e., (\ref{NT_BH}), with spin arrays. This culminates in an identification of a physical system, a spin array with a local anisotropy, that can be mapped into the model at hand.

\section{The number theoretic implementation of the Bose-Hubbard model}
\label{Section_NT}

Recall that the bosonic Fock space $\mathbb{H}^\odot$ is a separable Hilbert space built as follows: First, set the single-particle space $\mathbb{H}_{\text{SP}} = \text{span}\{|p\rangle:\, p \text{ prime}\}$.
Subsequently, let
\[
\mathbb{H}^\odot = \bigoplus\limits_{k=0}^\infty \mathbb{H}_{\text{SP}}^{\odot k}, \quad \mbox{ where }  \, \, \mathbb{H}_{\text{SP}}^{\odot 0} = \mathbb{C}.
\]
Here, $\odot$ signifies the symmetric tensor product. Note that the subspace $\mathbb{H}_{\text{SP}}^{\odot k}$ is spanned by vectors of the form
$
|p_1\rangle \odot \ldots \odot |p_k\rangle$,
where $p_1, \ldots p_k$ is any collection of $k$ primes, possibly with repetitions. Uniqueness of the prime decomposition of integers allows one to identify
\[
|p_1\rangle\odot  \ldots \odot |p_k\rangle = | n \rangle,
\]
where $n = p_1 \cdot \ldots \cdot p_k$ is the prime decomposition of $n$ (where possible repetitions are implicit). Thus, $\mathbb{H}^\odot = \mbox{ span }\{ |n\rangle: n \in \mathbb{N}\}.$

Next, consider the Hilbert space of square-summable arithmetic functions
\[
\ell_2(\mathbb{N})= \left\{f: \mathbb{N}\rightarrow \mathbb{C}: \|f\|^2=\sum_{n\in \mathbb{N}} \, |f(n)|^2 < \infty \right\}
\]
 with the standard inner product $\langle g | f \rangle = \sum_{n\in \mathbb{N}} \, g(n)^*f(n) $. It is clear that the set of point measures $\delta_n$, $n\in \mathbb{N}$ furnishes an orthonormal basis. Thus, square-integrable arithmetic functions can be represented in the distinguished basis as follows:
\begin{equation}\label{f_in_basis}
 f(x) = \sum_{n\in \mathbb{N}} \, f(n) \, \delta_n(x).
\end{equation}
Note that $f(n) = \langle \delta_n | f\rangle$.
It is now evident that $\ell_2(\mathbb{N})$ is naturally isomorphic with $\mathbb{H}^\odot$. Indeed, identifying the two sets of basis vectors via
\[
\delta_1 \mapsto 1, \quad \delta_{n} \mapsto |n\rangle
\]
and subsequently extending this map via linearity furnishes a unitary equivalence between the two spaces.

\subsection{The number-theoretic creation and annihilation operators}\label{subsec_a_ahat}

With this understood, we introduce the set of creation and annihilation operators that act on arithmetic functions. Consider the prime decomposition of an integer $n$ in the form
\[
 n = \prod_{p \in \mathcal{P}} \, p^{a_p(n)}.
\]
 This defines $a_p(n)$, i.e., the multiplicity of $p$ in the prime decomposition of $n$.
  It is convenient to retain two equivalent descriptions of operators: via their action on the distinguished basis, and via their action on functions. In every particular instance, the equivalence is made explicit by (\ref{f_in_basis}).  And so, for every prime $p$ we define the annihilation and creation operators as follows:
\begin{equation}\label{on_arithmetic}
  \begin{split}
    \hat{a}_p \, \delta_n = \sqrt{a_p(n)}\,\delta_{\frac{n}{p}} ,\quad \mbox{ or, equiv. } \quad & \hat{a}_{p}[f] (x)  = \sqrt{a_p(x)+1}\,f(x p)  \\
    & \\
     \hat{a}_p^\dagger \, \delta_n = \sqrt{a_p(n)+1}\,\delta_{np} ,\quad \mbox{ or, equiv. } \quad &  \hat{a}_{p}^\dagger [f] (x)   = \sqrt{a_p(x)}\, f\left(\frac{x}{p}\right).
   \end{split}
\end{equation}
Here and henceforth, we adopt the convention that whenever $n/p$ (resp. $x/p$) is not an integer, the expression $\delta_{n/p}$ (resp. $f(x/p)$) is replaced by zero.
A direct calculation shows that $\hat{a}, \hat{a}^\dagger$ satisfy the BCCR, (\ref{bCCR}).

As is standard, the number operators are defined as $\hat{N}_p = \hat{a}_p^\dagger \hat{a}_p$. A direct check yields $\hat{N}_p \delta_n = a_p(n)\delta_n$. The total particle number operator is $\hat{N} = \sum_{p\in \mathcal{P}} \hat{N}_p$. Recall the standard number-theoretic notation
\begin{equation}\label{Omega}
  \Omega(n) = \sum_{p} a_p(n).
\end{equation}
Later on we will also use a related arithmetic function defined as:
\begin{equation}\label{Q}
  Q(n) = \sum_{p} a_p(n)^2.
\end{equation}
Thus,
\begin{equation}\label{the_Ns}
  \hat{N} \delta_n = \sum_{p\in \mathcal{P}} \hat{N}_p\, \delta_n = \Omega(n)\, \delta_n, \quad
  \sum_{p\in \mathcal{P}} \hat{N}_p^2\, \delta_n = Q(n)\, \delta_n .
\end{equation}
It follows that
\begin{equation}\label{k-space}
\mathbb{H}_{\text{SP}}^{\odot k} = \mbox{ span }\{\delta_n: \Omega(n) = k \} = \{f : \hat{N} [f] = k\, f \},
\end{equation}
where we have utilized the identification $\mathbb{H}_{\text{SP}}^{\odot } \equiv \ell_2(\mathbb{N})$ to interpret $\mathbb{H}_{\text{SP}}^{\odot k} $ as a subspace in $\ell_2(\mathbb{N})$.
\vspace{.5cm}

\noindent
\emph{Example.} As mentioned above, the number-theoretic implementation of the creation and annihilation operators is fully equivalent with the standard one. We will illustrate the nature of this equivalence with an example: For simplicity, let us examine a model with just two sites, say, $\mathbb{H}_{\text{SP}} = \mbox{ span }\{\delta_2, \delta_3 \} $. In such a case,  $\mathbb{H}_{\text{SP}}^{\odot k} = \mbox{ span }\{\delta_{2^\alpha 3^{k-\alpha}}: \alpha = 0, 1, \ldots k \}$. Merely changing the notation, one obtains an alternative description $\mathbb{H}_{\text{SP}}^{\odot k} = \mbox{ span }\{|0\rangle |k\rangle, |1\rangle |k-1\rangle, \ldots |k\rangle |0\rangle,  \}$. In the latter notation, formulas (\ref{on_arithmetic}) reduce to the familiar expressions:
\[
  \hat{b}_1\, |j\rangle |k-j\rangle = \sqrt{j}\,\, |j-1\rangle |k-j\rangle, \quad
   \hat{b}_1^\dagger\, |j\rangle |k-j\rangle = \sqrt{j+1}\,\, |j+1\rangle |k-j\rangle,
\]
\[
  \hat{b}_2\, |j\rangle |k-j\rangle = \sqrt{k-j}\,\, |j\rangle |k-j-1\rangle, \quad
   \hat{b}_2^\dagger\, |j\rangle |k-j\rangle = \sqrt{k-j+1}\,\, |j\rangle |k-j+1\rangle,
\]
where we use $  \hat{b}_1,   \hat{b}_2$ to denote the effect of  $\hat{a}_2,   \hat{a}_3$ in the new notational convention, etc.

\begin{figure}[t]
\centering
\includegraphics[width=60mm]{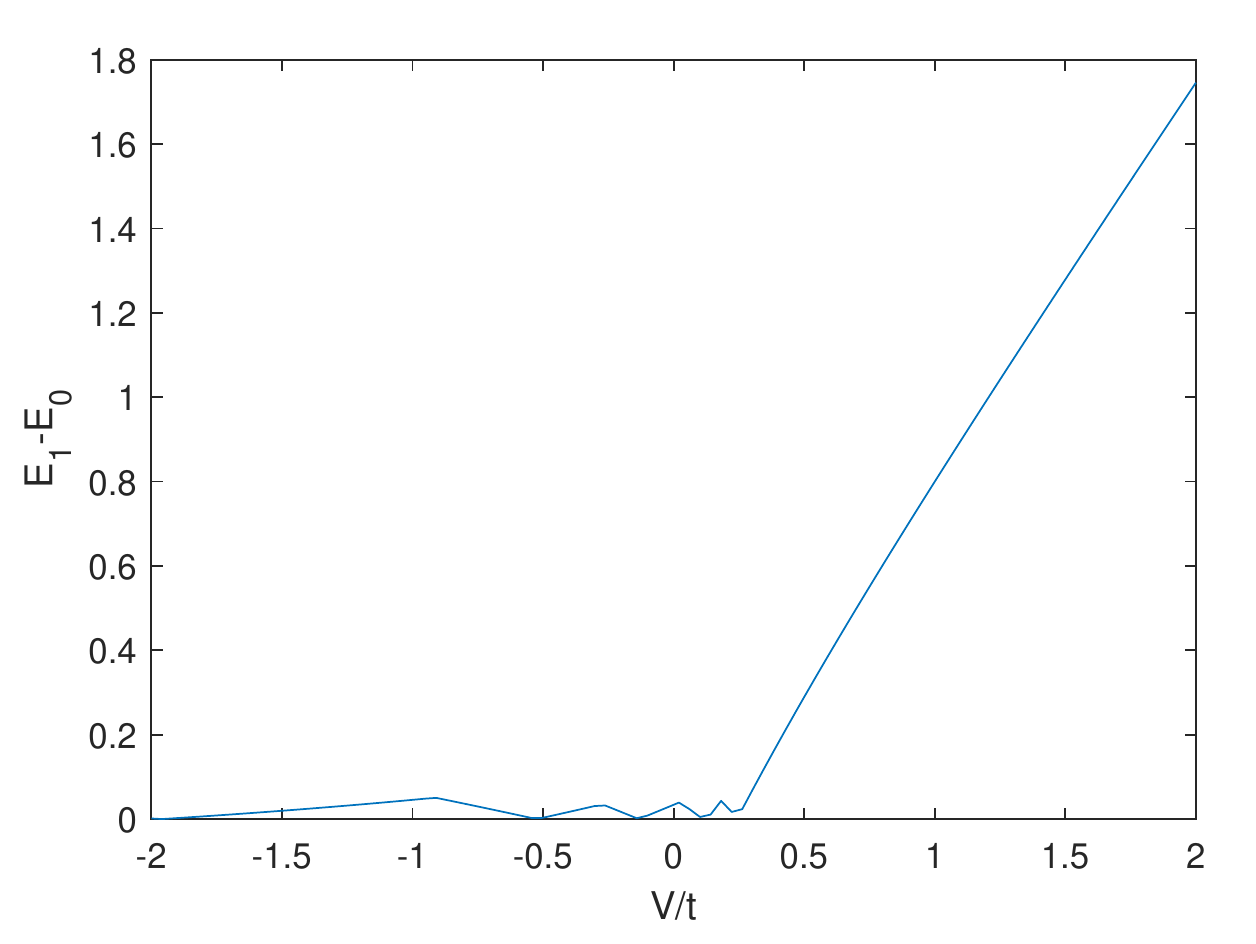}
\caption{As expected, in some instances the proposed method merely reproduces findings previously reported elsewhere, e.g. the above graph is qualitatively similar to one found in \cite{Jack_Yamashita}.  The figure displays dependence of the gap between the ground and the fist excited energy levels of the Bose-Hubbard Hamiltonian, i.e., $E_1- E_0$, as a function of $V/t$. Here, $t= -.1$, $\mu = 0$, and the size of the Hamiltonian matrix is $1024\times 1024$. }
\label{E1_minus_E0}
\end{figure}

Nevertheless, there are some advantages to the implementation (\ref{on_arithmetic}) over the standard one. First, in the numerical context, it facilitates computation incorporating multi-particle states with efficiency, e.g., a model based on a Hamiltonian matrix as small in size as $2^{10}\times 2^{10}$  already incorporates consistently many multi-particle states (with particle number of $10$ or less); it also incorporates states supported on up to four sites (because $2\times 3\times 5\times 7 = 210 < 1024$). It does not encompass all states from the direct sum of the respective multi-particle spaces, but a cross-section of that space together with an undistorted representation of the creation/annihilation operators in that cross-section. These characteristics of the model are essential in assessing the phase transitions, as discussed in Section \ref{Section_numerical}. In some other investigations they do not seem to play a big role, e.g. the graph in Fig. \ref{E1_minus_E0} merely confirms a result in \cite{Jack_Yamashita}, which has been obtained via the standard implementation of the model.

Second, crucially, the number theoretic point of view on the fully infinite-dimensional chain of sites facilitates the application of harmonic analysis on the multiplicative group of positive rationals, leading to new insights into the physics of the Bose-Hubbard model, see Sections \ref{Section_FT_n_Kast}-\ref{sec-physical}.

\subsection{Invariance of k-particle spaces}

An inspection of formulas (\ref{on_arithmetic}) leads to an observation that spaces $\mathbb{H}_{\text{SP}}^{\odot k}$ are invariant for the Hamiltonian (\ref{NT_BH}), i.e.
\begin{equation}\label{Block_factorization}
   \mathcal{H}: \mathbb{H}_{\text{SP}}^{\odot k} \rightarrow \mathbb{H}_{\text{SP}}^{\odot k},
\end{equation}
so that
\begin{equation}\label{H_N_commute}
  [\mathcal{H}, \hat{N} ] =0.
\end{equation}
Consider the matrix of  $\mathcal{H}$ restricted to $\mathbb{H}_{\text{SP}}^{\odot k}$ in the distinguished basis $\{\delta_n: \, \Omega(n) = k\} $, with basis elements ordered by increasing $n$. When $k>1$ these matrices have a unique peculiar structure. However, for $k=1$ the matrix has the form of a Toeplitz operator:
\begin{equation}\label{Toep}
\left(
  \begin{array}{cccc}
    -\mu & -t &  & \\
    -t & -\mu & -t &  \\
     & -t & -\mu & -t \\
     &  &  & \ddots  \\
  \end{array}
\right)
\end{equation}

It is well known that the spectrum of this operator is the interval $[-\mu - 2t,  - \mu + 2t]$. \emph{A fortiori}, the spectrum of $\mathcal{H}$ contains this interval.  Moreover, the spectral points correspond to sequences of approximate eigenfunctions. In the absence of disorder (i.e., when $t=0$), the interval degenerates to the point $-\mu$, which is then an eigenvalue.

\subsection{The sparse structure of the Bose-Hubbard Hamiltonian}

We now turn to the analysis of the Hamiltonian. Apart from the full infinite-dimensional Hamiltonian (\ref{NT_BH}), it is interesting to consider finite-dimensional implementations which arise from restricting $\ell_2(\mathbb{N})$ to the subspace
\begin{equation}\label{subspace_F_N}
 F_N = \mbox{ span } \{ \delta_n:  n \leq N \}.
\end{equation}
 Note that while $F_N$ are finite-dimensional, they contain many multi-particle states. This will enable us to discern nontrivial quantum states in the dynamic of the corresponding $N$-by-$N$ Bose-Hubbard Hamiltonian $\mathcal{H}|_{F_N}$.
All direct calculations are merely identical in the finite-dimensional and infinite-dimensional cases. It is easy to adapt general calculations to either case, simply by a proper interpretation.

Let us consider the problem of solving the stationary Schr\"{o}dinger equation:
\begin{equation}\label{Schrod}
  \left(\mathcal{H}_{\mbox{order}} -t\,\mathcal{H}_{\mbox{hop}}  \right) [f] = Ef,
\end{equation}
where we have abbreviated as $\mathcal{H}_{\mbox{order}}$ the $t=0$ part of the Hamiltonian, and as $\mathcal{H}_{\mbox{hop}} $ the  (hopping) part that is controlled by $t$.
In the distinguished basis $f = \sum_{n} z_n \delta_n$ with $z_n \in \mathbb{C}$ satisfying $\sum |z_n|^2 =1$.  The order component $\mathcal{H}_{\mbox{order}}$ is diagonal in this basis; in fact
\begin{equation}\label{diag_BH}
 \langle \delta_n \, | \, \mathcal{H}_{\mbox{order}} [f] \rangle =  \left[\frac{U}{2} Q(n) - \left( \frac{U}{2} + \mu\right)\Omega(n)\right] z_n ,
\end{equation}
which follows directly from (\ref{the_Ns}) with definitions (\ref{Omega}), (\ref{Q}).
The hopping part is off-diagonal. Namely,
\begin{align}\label{hopping}
  \langle \delta_m\, |& \,  \mathcal{H}_{\mbox{hop}} [f] \rangle  = \nonumber\\
  & \sum_{n} \sqrt{a_{p_{n+1}}(m) +1 }\sqrt{a_{p_n}(m)}\, z_{mp_{n+1}/p_n}
  + \sqrt{a_{p_{n}}(m) +1 }\sqrt{a_{p_{n+1}}(m)}\, z_{mp_{n}/p_{n+1}}.
  \end{align}
\vspace{.5cm}

\noindent
\emph{Example.} It is interesting to list the explicit result of (\ref{hopping}) for a few values of $m$:
\[
\begin{tabular}{|l|r|}
  \hline
  m &  $\langle \delta_m\, | \,  \mathcal{H}_{\mbox{hop}} [f] \rangle$ \\
  \hline
  1 & $0$ \\
  2 & $z_3$ \\
  3 & $z_2 + z_5$ \\
  4 & $\sqrt{2} \, z_6 $ \\
  5 & $z_3 + z_7$ \\
  6 & $\sqrt{2} (z_4+z_9) + z_{10}$ \\
  7 & $z_5 + z_{11}$ \\
  8 & $\sqrt{3} z_{12}$ \\
  \hline
\end{tabular}
\]
In the case of larger matrices one may observe that the non-diagonal entries tend to concentrate along centrally convergent rays. It follows from formulas  (\ref{diag_BH}-\ref{hopping}) that the number of nonzero entries in the n-th row is between $\omega (n)+1$ and $2\omega(n) +1$, where $\omega(n)$ denotes the number of distinct prime factors of $n$. Based on the known properties of the summatory function of $\omega$, this implies that \emph{asymptotically} the matrix of size $N$ contains $O(N\log \log N )$ nonzero entries.

\section{Manifestations of phase transitions}
\label{Section_numerical}
Consider a restriction of $\mathcal{H}|_{F_N}$ to the subspace $F_N \cap \mathbb{H}_{\text{SP}}^{\odot 1}$. Representing the operator in the basis $\{\delta_p: p <N \}$ with primes $p$ in the natural order, one obtains a truncation of matrix  (\ref{Toep}) of size $\pi(N)$ equal to the number of primes not exceeding $N$. The spectrum of such a matrix is well-known: it consists of points
\[
E_{N,k} =  - \mu -2t\cos\left( \frac{k}{\pi(N)+1}\pi \right): \quad k = 1, 2,\ldots \pi(N).
\]
Note that these discrete points fill the interval $[-\mu - 2t, -\mu +2t]$ more and more densely as $N$ increases. This is important partial information about the spectrum of $\mathcal{H}|_{F_N}$. We are not aware of a method to find a close-form formula for the general Hamiltonian. However, the number-theoretic implementation enables a very satisfactory computer-assisted treatment of the finite-dimensional reduction.

\subsection{The dependence of the qualities of the ground state on the model's parameters}

\begin{figure}[t]
\centering
\includegraphics[width=120mm]{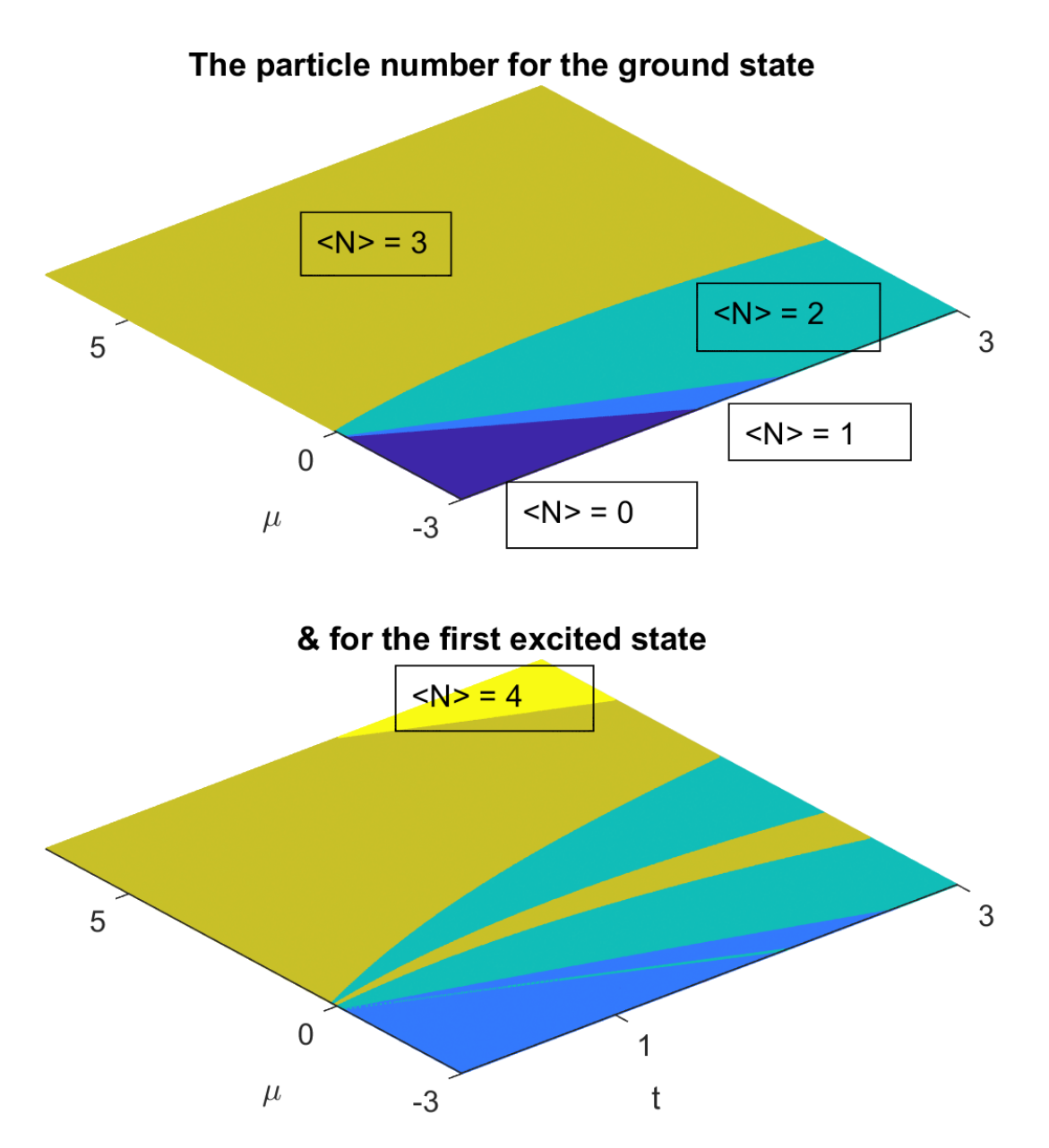}
\caption{The phase diagrams of $\langle \hat{N} \rangle $ of the ground states (top) and for the first excited states (bottom) at $U = 10$; the computation utilized Hamiltonian matrices of size $150\times 150$. The color coding is consistent in both diagrams, marking regions with $\langle \hat{N} \rangle = 0,1, 2, 3, 4$.
 }
\label{Ground_n_Excited_phases}
\end{figure}

The number-theoretic setting of the Hamiltonian $\mathcal{H}|_{F_N}$ lends itself to high-accuracy computer simulation. Indeed, clearly, formulas (\ref{on_arithmetic}) can be implemented without any error, apart from approximations for the irrational numbers. This stands in stark contrast to the practice of expressing the creation and annihilation operators via discretized differential operators, or finite-dimensional approximations of essentially infinite matrices (necessarily violating the BCCR).   The algorithm to construct the Hamiltonian matrix requires prime factorization of all integers $n= 2, 3, 4, \ldots N$. Hence it is not efficient and prohibitively costly for extremely large values of $N$. Nevertheless, for relatively small $N$ computation of eigenvalues and eigenvectors of the Hamiltonian yields results with an essentially perfect accuracy.
One of the first observations is the occurrence of the dependence of the quality of the ground states and first few excited states on the value of parameters. The diagram in Fig. \ref{Ground_n_Excited_phases} illustrates that for some values of the parameters the low lying states are superpositions of single-particle states, whereas for other this is no longer true as the superpositions involve multi-particle states. Note that this fact is rigorous, i.e. it is not burdened with a numerical approximation error.
The existence of various phases, marked by distinct values of $\langle \hat{N} \rangle$ is made evident by the figures. It is demonstrated in the next section that the phases of the structure identified here remain unchanged under certain natural continuous unitary deformations; for a summary see Subsection \ref{Subsection_flow}. Again, such deformations are easily interpreted in both the finite- and the infinite-dimensional cases.

\subsection{Phase transitions in the grand canonical ensemble}

It is worthwhile to begin the discussion with a few observations. First, in the finite-dimensional space $F_N$ we have
\[
\mbox{trace}\, \mathcal{H}|_{F_N} =  \mbox{trace} \, \mathcal{H}_{\mbox{order}}|_{F_N} = \frac{U}{2} \, \sum_{n \leq N} Q(n) -\left( \frac{U}{2} + \mu\right)\, \sum_{n \leq N}\Omega(n) .
\]
This is a linear function of $U$ and $\mu$ with the coefficients that are number-theoretic constants. Similarly, we have
\[
\mbox{trace}\, \hat{N}|_{F_N} = \sum_{k=1}^\infty k \cdot \dim F_N \cap \mathbb{H}_{\text{SP}}^{\odot k}.
\]
The sum is in fact finite as $ \dim F_N \cap \mathbb{H}_{\text{SP}}^{\odot k}=0$ once $k$ exceeds the highest possible complexity of a number less or equal than $N$. Thus, $\mbox{trace}\, \hat{N}|_{F_N}$ is a number-theoretic constant independent of the model's parameters.

\begin{figure}[t]
\centering
\includegraphics[width=120mm]{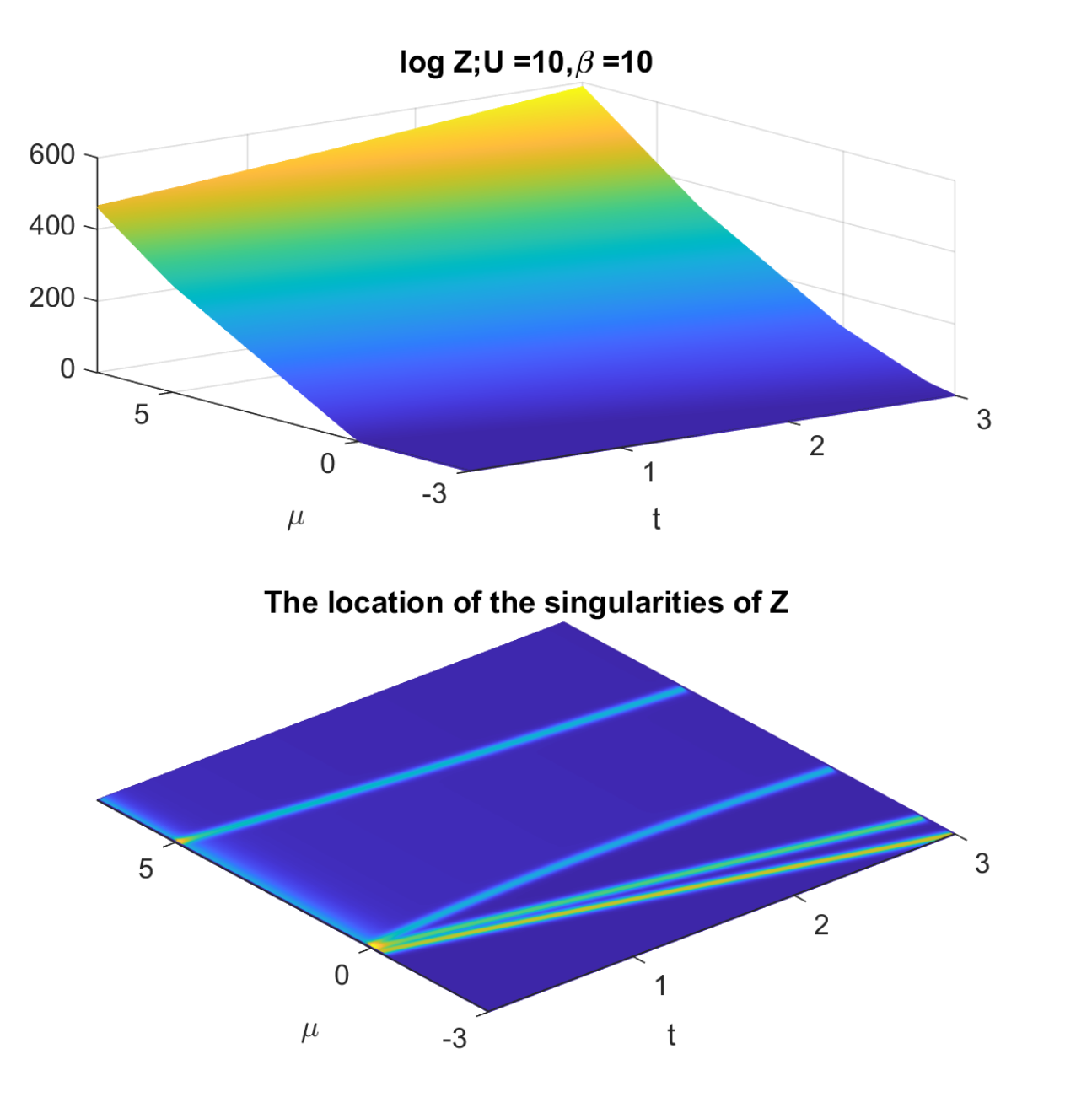}
\caption{The graph of $\log Z(\mu,t)$ for fixed $U=10$, $\beta = 10$ (in units such that $k_B =1$); the computation was based on Hamiltonian matrices of size $150\times 150$. The lower graph displays the location of singularities of the surface. The singularities have been detected by applying the discrete Laplacian filter. They align along four lines which partition the area shown into five distinct quantum-phase regions. The singular line closest to the viewer is clearly manifested as a crease in the graph of $\log Z$, while all other are harder to discern with the naked eye.
}
\label{Z_phases}
\end{figure}

At the same time, the grand canonical ensemble partition function $Z$ depends on the model parameters as well as on the temperature $T$, i.e. $Z = Z(\mu, t, U, T)$ . It is given, \cite{Reichl} (Section 6.2), via the closed-form formula:
\begin{equation}\label{Z_grand_canonical}
  Z = \mbox{trace}\, \exp \left[-\beta\,(\mathcal{H}|_{F_N} - \mu\,  \hat{N}|_{F_N})\right],
\end{equation}
where $\beta = 1/k_B T$ and $k_B$ is the Boltzmann constant.
Fig. \ref{Z_phases} (upper part) displays the graph of $\log Z(\mu, t)$ in the rectangle $(\mu, t)\in [-3,7]\times [.1, 3]$. There is a clearly visible crease in the graph, but also a few less pronounced creases, indiscernible to the naked eye. These creases are detected via an application of the discrete Laplacian filter. In other words the matrix containing the values of $\log Z$ is convolved with the matrix
\[
\left(
  \begin{array}{ccc}
    0 & 1 & 0 \\
    1 & -4 & 1 \\
    0 & 1 & 0 \\
  \end{array}
\right)
\]
and, subsequently, the artifact created by the edges is trimmed. The resulting matrix approximates, up to scale, $\Delta \log Z(\mu, t)$ with $\Delta = \partial_\mu^2 + \partial_t^2$. The lower part of Fig. \ref{Z_phases} displays the color-map graph of this matrix. This surface brings out the singularities of $\log Z(\mu, t)$, which clearly divide the displayed square into five regions, i.e., five distinct quantum phases. The graph has been computed exploring Hamiltonian matrices of size $150\times 150$. This matters, e.g., a similar computation based on a smaller matrix does not capture the farthest singular line. On the other hand, increasing the size of the matrix, even significantly, does not result in any new information in this region of the $(\mu, t)$ plane. Exploration of farther regions in the plane will require higher computational resources, and is not undertaken in this article.

\section{The Fourier-dual point of view}
\label{Section_FT_n_Kast}

We will demonstrate that the creation and annihilation operators (\ref{on_arithmetic}) and the Fock space in its $\ell_2(\mathbb{N})$ representation admit a type of Fourier-dual representation. To this end we need to briefly outline Harmonic analysis on the group of positive rationals. Outside the theoretical framework of the Pontryagin duality the latter is rather little known, albeit it has been applied in pioneering ways in the Analytic Number Theory, \cite{Elliott}. For the reader's convenience we give a brief outline of the foundations and basic features of this technique.

\subsection{Harmonic analysis on the multiplicative group of positive rationals }

First note the isomorphism of abelian groups:
\begin{equation}\label{rationals}
 \mathbb{Q}_+ \equiv \bigoplus\limits_{p: \in \mathcal{P}} \mathbb{Z} \quad \mbox{ given by the prime factorization  } \quad \mathbb{Q}_+ \ni w = \prod_{p\in \mathcal{P}} p^{a_p}, \, a_p \in\mathbb{Z}.
\end{equation}
In light of this, the dual group of $\mathbb{Q}_+$ is
\[
\hat{\mathbb{Q}}_+ = \prod_{p \in \mathcal{P}} U(1).
\]
When equipped with the product topology $\hat{\mathbb{Q}}_+$ is, by virtue of the Tychonoff Theorem, a compact space. Secondly, it admits a unique Borrel measure $d\vec{\mu}$, which satisfies
\[
d\vec{\mu} \left( (\alpha_2, \beta_2]  \times  \ldots \times (\alpha_p, \beta_p] \times (0,1]  \times (0,1] \times \ldots  \right) =
|\beta_2 - \alpha_2| \ldots |\beta_p - \alpha_p|.
\]
In particular, $d\vec{\mu} ( \hat{\mathbb{Q}}_+ ) =1 $, i.e. the measure is probabilistic. At the same time, $\mathbb{Q}_+$ itself is equipped with the discrete (counting) measure. It is often  useful to identify $\hat{\mathbb{Q}}_+$ with the set of completely multiplicative functions $\chi: \mathbb{Q}_+\rightarrow U(1)$, i.e. functions that satisfy $\chi(uw)= \chi(u)\chi(w)$. Namely,
\begin{equation}\label{mult_chi}
\prod_{p \in P} U(1) \ni (\theta_2, \theta_3, \theta_5, \ldots ) \,\, \mbox{ corresponds to } \chi \mbox{ characterized by }\,\, \chi(p) = p^{i \theta_p} , p \in \mathcal{P}.
\end{equation}
It is also useful to introduce a change of variable setting $ \mu_p = \theta_p\,\log p/(2\pi)$,
so that
\[
\frac{\log p}{2\pi}\int\limits_{0}^{2\pi/\log p } p^{ik\theta_p} \, d\theta_p =
\int\limits_{0}^{1} e^{2\pi ik\mu_p} \, d\mu_p = \left\{\begin{array}{cc}
                                                         1 & k = 0 \\
                                                         0 & k \neq 0
                                                       \end{array}\right. .
\]
Utilizing these identifications---identifying $\chi$ with  $(\theta_2, \theta_3, \theta_5, \ldots )$ and with $(\mu_2, \mu_3, \mu_5, \ldots )$----the Fourier transform appears in several different guises. First, it is defined via:
 \begin{equation}\label{FT_def1}
   \hat{f}(\chi) = \sum_{w\in \mathbb{Q}_+} f(w)\,\chi(w)^* \mbox{ where } f:\mathbb{Q}_+ \rightarrow \mathbb{C}.
 \end{equation}
This defines $\hat{f}: \hat{\mathbb{Q}}_+ \rightarrow \mathbb{C}$ with $\chi$ as its argument. The inverse transform is then given by
\begin{equation}\label{FT_Inv_def1}
  f(w) = \int \hat{f}(\chi)\, \chi(w)\, d\vec{\mu}(\chi) = \int\limits_{0}^{1} \int\limits_{0}^{1} \int\limits_{0}^{1} \ldots \hat{f}(\chi)\, \chi(w) \, d\mu_2\, d\mu_3\, d\mu_5\ldots
\end{equation}
This  notation is useful in particular when expressing the following fundamental properties: First, for a fixed arbitrary $u \in \mathbb{Q}_+ $, a direct calculation shows
\begin{equation}\label{mult}
 \mbox{Let } g(w): = f(uw) \mbox{ for all } w \in \mathbb{Q}_+.  \mbox{ Then } \hat{g}(\chi) = \chi(u^{-1}) \hat{f}(\chi).
\end{equation}
Second, note that the measure $d\mu(\chi)$ is invariant with regards to circular shifts along the $U(1)$ components. Thus, for a fixed collection $\vec{\nu} = (\nu_2, \nu_3, \nu_5, \ldots)$ if we define $\hat{g}$ via
\[
\hat{g}(\vec{\mu} ) : = \hat{f}(\vec{\mu} - \vec{\nu} ),
\]
then $\|\hat{f}\| = \| \hat{g} \|$ since $d\vec{\mu}$ is shift invariant. Calculating the inverse transform we readily obtain
\begin{equation}\label{shift}
  g(w) = \chi_{\vec{\nu}}(w) \, f(w), \quad \mbox{ where } \quad \chi_{\vec{\nu}} \equiv (\nu_2, \nu_3, \nu_5, \ldots).
\end{equation}

We now turn attention to the Hilbert-space theoretic aspects. In some ways it is more akin to that of $L_2(\mathbb{R})$, rather than $L_2(U(1))$. That is because the characters are not square integrable functions, indeed:
\[
\|\chi\|^2 =  \sum_{w\in \mathbb{Q}_+} \chi(w)\, \chi(w)^* =  \sum_{w\in \mathbb{Q}_+} 1 = \infty .
\]
We will briefly discuss the Parseval identity. On one hand, it follows from the general Pontryagin Theorem. On the other hand, it is instructive to observe it directly in this context. The calculation becomes more explicit with the use of (\ref{rationals}) to identify rational numbers with finitely supported sequences of integers:
\[
\vec{a} = (a_2, a_3, a_5, \ldots) \in \bigoplus\limits_{p: \in P} \mathbb{Z}.
\]
Accordingly, we use the notation:
$
\vec{a}\cdot \vec{\mu} = a_2\mu_2 + a_3\mu_3 + a_5\mu_5 + \ldots \in \mathbb{C}.
$
We can express the Fourier transform of the point measure:
\[
\delta_{\vec{a}} (x ) \mapsto e^{2\pi i \vec{a}\cdot \vec{\mu}} ,
\]
so that $x$ and $\hat{\mu}$ are dual variables.
Thus, for an arbitrary function $f$ on $\mathbb{Q}_+$ we have
\begin{equation}\label{FT_def2}
  f(w) = \sum_{\vec{a}} f(\vec{a}) \, \delta_{\vec{a}} (w ) \quad \mapsto \quad \hat{f}(\vec{\mu}) = \sum_{\vec{a}} f(\vec{a}) \, e^{2\pi i \vec{a}\cdot \vec{\mu}} ,
\end{equation}
where the summation is over all $\vec{a} \in \bigoplus\limits_{p: \in P} \mathbb{Z}$. It is easy to see that definitions (\ref{FT_def1}) and (\ref{FT_def2}) are equivalent.
Now, suppose $f: \mathbb{Q}_+ \rightarrow \mathbb{C}$ is square summable, i.e.
\begin{equation}\label{l2_Q_+}
\|f\|^2 = \sum_{w\in \mathbb{Q}_+} |f(w)|^2 = \sum\limits_{\vec{a}}\,
| f(\vec{a})|^2 < \infty .
\end{equation}
Note that
\begin{equation}\label{intexp}
  \int d\vec{\mu} \,\, e^{2\pi i \vec{a}\cdot \vec{\mu}} = \left\{ \begin{array}{cc}
                                                                  1 & \vec{a} = 0 \\
                                                                  0 & \mbox{ oth.}
                                                                \end{array}
  \right.
\end{equation}
Next,  we examine the norm of $\hat{f}$:
\begin{align}
\|\hat{f}\|^2  &= \int d\vec{\mu}\,\, |\hat{f}(\vec{\mu} )|^2 =  \int d\vec{\mu}\, \left|\, \sum_{\vec{a}} f(\vec{a}) \, e^{2\pi i \vec{a}\cdot \vec{\mu}}\, \right|^2 \nonumber \\
   &  \nonumber\\
   &=  \int d\vec{\mu}\,\, \sum\limits_{\vec{a}}\, \left|
f(\vec{a} )\, \right|^2 + 2 \int d\vec{\mu}\,\, \sum\limits_{\vec{a}}\,f(\vec{a})\,\sum\limits_{\vec{b}\neq \vec{a}}\,\,
 f(\vec{b} )^*\, \exp [ - 2\pi i (\vec{a}-\vec{b})\cdot \vec{\mu}\, ] \nonumber \\
& \nonumber\\
&= \|f\|^2.\nonumber
\end{align}
Indeed,  the second term of the sum vanishes by virtue of (\ref{intexp}). That becomes evident when the order of summation and integration is reversed, which is allowed as summability is absolute.
This is equivalent to stating that as $\vec{a}$ runs over all finitely supported sequences of integers $\delta_{\vec{a}}(x)$ furnish an orthonormal basis in $\ell_2(\mathbb{Q}_+)$ while $\exp(2\pi i \vec{a}\cdot \vec{\mu})$ furnish such a basis in $L_2(\hat{\mathbb{Q}}_+, d\vec{\mu})$, so that the map (\ref{FT_def2}) is unitary.

\subsection{The Kastrup model of the creation and annihilation operators as the dual-equivalent to the arithmetic model}

An interesting model for a quantum harmonic oscillator, identified in \cite{Kastrup}, is related to the Hecke space $H_2(U(1))$, which consists of square integrable functions whose Fourier series do not admit any negative frequencies. The standard unitary basis consists of functions $e^{2\pi i n \mu_p}$ with $n = 0,1,2,\ldots $. Note that the Fourier transform introduced in the previous section establishes an equivalence of the spaces and subspaces as follows:
\begin{equation}\label{diagram}
\begin{array}{ccc}
  \bigotimes\limits_{p\in\mathcal{P}} H_2(U(1)) & \subset & \bigotimes\limits_{p\in\mathcal{P}} L_2(U(1))\\
  \updownarrow\mbox{FT} & & \updownarrow\mbox{FT} \\
  \ell_2(\mathbb{N}) & \subset & \ell_2(\mathbb{Q}_+)
\end{array}
\end{equation}
Even though the arithmetic model of the Fock space corresponds to the Hecke subspace, the group-duality based theory requires that in order to understand the whole picture we cannot loose the sight of the entire $ \ell_2(\mathbb{Q}_+)$.

Next, in order to define the quantum harmonic oscillator (for each $p$) one starts with the following three fundamental operators:
\begin{equation}\label{circ_observ}
\begin{split}
    &  K_{0,p} =  \frac{1}{2\pi i}\partial_{\mu_p} + \frac{1}{2},  \\
  & K_{1,p} = \cos (2\pi\mu_p) \frac{1}{2\pi i}\partial_{\mu_p} + \frac{1}{2}e^{2\pi i \mu_p}, \\
  & K_{2,p} = \sin (2\pi\mu_p) \frac{1}{2\pi i}\partial_{\mu_p} + \frac{1}{2i}e^{2\pi i \mu_p}.
  \end{split}
\end{equation}
Note that elements of the space $H_2(U(1))$ are functions of the form $f = \sum_{n=0}^{\infty} f_n e^{2\pi i n \mu}$ which can be identified with square-summable sequences $(f_0, f_1, \ldots )$. A multiplication operator  $e^{2\pi i \mu}$ acts as a right shift $(f_0, f_1, \ldots ) \mapsto (0, f_0, f_1, \ldots )$. Its adjoint is the multiplication by $e^{-2\pi i \mu}$, i.e. a left-shift $(f_0, f_1, \ldots ) \mapsto (f_1, f_2, \ldots )$. We have $\cos (2\pi \mu) = (e^{2\pi i \mu} + e^{-2\pi i \mu})/2$, etc. With this understood, it is easily verified that all operators $K_{0,p}$, $K_{1,p}$, and $K_{2,p}$  are self-adjoint. Indeed, we have,
\begin{equation}\label{Ks_coord}
\begin{split}
   K_0 (f_0, f_1, f_2 \ldots ) = & \left(\frac{1}{2} f_0, \frac{3}{2} f_1, \frac{5}{2} f_2 \ldots \right) \\
    K_1 (f_0, f_1, f_2 \ldots ) = & \left(\frac{1}{2} f_1, \frac{1}{2} f_0 + f_2, f_1 + \frac{3}{2} f_3, \frac{3}{2} f_2 + 2f_4, \ldots \right) \\
   K_2 (f_0, f_1, f_2 \ldots ) = & \frac{1}{i}\left(-\frac{1}{2} f_1, \frac{1}{2} f_0 - f_2, f_1 - \frac{3}{2} f_3, \frac{3}{2} f_2 - 2f_4, \ldots \right)
\end{split}
\end{equation}
 (We have suppressed the index $p$ as the coordinate representations of these operators are the same for all $p$.)  The following commutation relations are also verified via direct calculation:
\begin{equation}\label{commutation}
  [\,K_{0,p}, K_{1,p} \,] = i K_{2,p},\quad   [\,K_{0,p}, K_{2,p} \,] = -i K_{1,p},\quad   [\,K_{1,p}, K_{2,p} \,] = -i K_{0,p}.
\end{equation}
Moreover, these observables are fundamental to the system, i.e. the creation and annihilation operators may be reconstructed from them. Namely, first define
\[
\begin{split}
   K_{+,p} = & K_{1,p} + i K_{2,p} = e^{2\pi i \mu_p} \left( \frac{1}{2\pi i} \partial_{\mu_p} + 1 \right), \\
    K_{-,p} =  & K_{1,p} - i K_{2,p} = e^{-2\pi i \mu_p}  \frac{1}{2\pi i} \partial_{\mu_p}.
\end{split}
\]
It follows that $ K_{+,p}^\dagger = K_{-,p}$. Indeed, in the standard basis, these operators assume the form:
\begin{equation}\label{pmcoord}
\begin{split}
   K_+ (f_0, f_1, f_2 \ldots ) = &  \left(0, f_0, 2 f_1, 3f_2, \ldots \right) \\
   K_- (f_0, f_1, f_2, f_3 \ldots ) =& \left(f_1, 2 f_2, 3f_3, \ldots \right)
\end{split}
\end{equation}
Second, one defines:
\begin{equation}\label{creat-annihil}
  \hat{a}_p = (K_{0,p} + \frac{1}{2})^{-1/2}\, K_{-,p},\quad \hat{a}_p^\dagger = K_{+,p}\, (K_{0,p} + \frac{1}{2})^{-1/2} .
\end{equation}
These are the creation and annihilation operators, which satisfy the bosonic canonical commutation relations:
\[
[\, \hat{a}_p, \hat{a}_q^\dagger \, ] = \delta_{p,q} .
\]
Indeed, in the standard basis these operators assume the standard form
\[
\hat{a} (f_0, f_1, f_2, f_3, \ldots ) = \left( f_1, \sqrt{2} f_2, \sqrt{3} f_3, \ldots \right)
\]
\[
\hat{a}^\dagger (f_0, f_1, f_2 \ldots ) = \left( 0, f_0, \sqrt{2} f_1, \sqrt{3} f_2, \ldots \right)
\]
All the operators listed above act on the functions $\hat{f}(\vec{\mu})$. The Fourier transform makes it possible to express these as acting on the arithmetic functions $f= f(\vec{a})$ or, equivalently, $f = f(w)$. By abuse of notation we will retain the same symbols for the Fourier transformed operators. Recall that FT establishes one-to-one correspondence between $\delta_{\vec{a}}$ and $\exp (2\pi i \vec{a}\cdot \vec{\mu})$. Using this one translates (\ref{Ks_coord}) and (\ref{pmcoord}) into:
\begin{equation}\label{on_sum_Z}
  \begin{split}
     K_{0,p}\,\delta_{\vec{a}}  = (a_p+\frac{1}{2})\, \delta_{\vec{a}},\quad \mbox{ or, equiv. } & K_{0,p}[f] (w)  = (a_p(w)+\frac{1}{2})\, f(w),\\
        K_{+,p}\,\delta_{\vec{a}}  = (a_p+1)\, \delta_{\vec{a} + e_p},\quad  \mbox{ or, equiv. } &  K_{+,p}[f] (w)  = (a_p(w)+1)\, f\left(\frac{w}{p}\right),\\
      K_{-,p}\,\delta_{\vec{a}}  = a_p\, \delta_{\vec{a}-e_p} , \quad \mbox{ or, equiv. }& K_{-,p}[f] (w)  = a_p(w)\,  f(wp).
  \end{split}
\end{equation}
It is understood that functions resulting from application of operators turn to zero on non-integer values. In the same way, one obtains (\ref{on_arithmetic}) (trading $w$ for $n$ to denote natural numbers). This closes our description in both Fourier dual settings of the fundamental structure of the Fock space with the set of creation and annihilation operators indexed by $\mathcal{P}$.
\vspace{.5cm}

\subsection{Unitary flows on the Fock space}\label{Subsection_flow}

We briefly point out another observation that stems from the dual picture as summarized in the diagram (\ref{diagram}). Namely, the flow on $\hat{\mathbb{Q}}_+$ given by $\vec{\mu} \mapsto \vec{\mu} - \tau\vec{\nu} $, where $\tau \in \mathbb{R}_+$ is the time parameter, defines a family of unitary automorphisms:
\begin{equation}\label{U_tau_flow}
 \sigma_\tau : \ell_2(\mathbb{N}) \rightarrow  \ell_2(\mathbb{N}).
\end{equation}
Indeed, (\ref{FT_def2}) shows that the effect of $\sigma_\tau$ on the Fourier coefficients is
\[
f(\vec{a})\mapsto f(\vec{a}) \exp(-2\pi i \,\tau\, \vec{a}\cdot\vec{\nu}).
\]
Evidently, the subspace $H_2(U(1))$ remains invariant, and so does the Fock space $\ell_2(\mathbb{N})$. Note also that the matrix of $\sigma_\tau$ is diagonal in the standard basis $\delta_n$. Indeed, using (\ref{shift}) and (\ref{mult_chi}), we obtain
\begin{equation}\label{shift_flow}
  \sigma_\tau: f(n) \mapsto \chi_{\tau\, \vec{\nu}}(n) \, f(n), \quad \mbox{ or, equiv.}\quad
  f(n) \mapsto \prod\limits_{p\in \mathcal{P}} p^{i\, a_p(n)\,\theta_p \tau} \, f(n).
\end{equation}
Consider the action of the flow on the Bose-Hubbard Hamiltonian (\ref{NT_BH}), i.e. $\tau \mapsto \sigma_\tau^\dagger \, \mathcal{H} \, \sigma_\tau=: \mathcal{H}_\tau$. Since the matrices $\sigma_\tau$ are diagonal, the sparse structure of the Hamiltonian remains fixed, only the phases of the off-diagonal entries drift with the flow of time.
As indicated in (\ref{mult_chi}), a multiplicative function is fully determined by its values on primes. Equivalently, the automorphisms $\sigma_\tau$ are determined by their action in the single-particle subspace of the Fock space. In particular, every $m$-particle subspace is invariant under the action of the flow. Thus, the different phases identified numerically in Section \ref{Section_numerical} remain invariant under the action of the flow.  In particular, Fig. \ref{Ground_n_Excited_phases} remains valid for all $\mathcal{H}_\tau$.

\section{A physical system that can be mapped into the model}
\label{sec-physical}

A type of physical model that can be mapped onto the Bose-Hubbard model discussed in the beginning of this text, is provided by a one-dimensional array of spin moments ${\bf S}_m$. Here, we consider nearest neighboring spins to be coupled through a Heisenberg interaction $J$ along with a local anisotropy field $D$, such that,
\begin{align}
{\cal H}=&
	-\sum_m
	\Bigl(
		J{\bf S}_m\cdot{\bf S}_{m+1}
		+
		D(S_m^z)^2
	\Bigr)
	.
\end{align}
Here, we point out that the spin operator ${\bf S}=(S_x,S_y,S_z)$ obeys the commutation relation $[S_i,S_j]=i\epsilon_{ijk}S_k$, where $\epsilon_{ijk}$ is the fully antisymmetric Levi-Cevita tensor. These relations are captured in (\ref{commutation}), hence, providing a direct link between the physical model and the number theoretic implementation of the Bose-Hubbard model.

For a ferromagnetic interaction ($J>0$) and uniaxial anisotropy ($D>0$), the excitations of the ferromagnetic ground state can be considered in terms of the Holstein-Primakoff expansion, that is,
\begin{subequations}
\begin{align}
S^z=&
	S-\hat{N}
	=
	S-\hat{a}^\dagger \hat{a}
	,
\\
S^+=&
	\sqrt{2S\biggl(1-\frac{\hat{N}}{2S}\biggr)}\,\hat{a}
	,
\\
S^-=&
	\hat{a}^\dagger\sqrt{2S\biggl(1-\frac{\hat{N}}{2S}\biggr)}
	,
\end{align}
\end{subequations}
where $S$ is the amplitude of the local spin moment.
It is worthwhile to remark that inverting (\ref{creat-annihil}), leads to the relations
\begin{equation}\label{creat-annihil_inverse}
 K_{-,p} = (\hat{N}_p+1)^{1/2} \,\hat{a}_p ,\quad
  K_{+,p} = \hat{a}_p^\dagger \,(\hat{N}_p+1)^{1/2},
\end{equation}
which constitute a form of the Holstein-Primakoff transform.

Retaining even orders up to quartic, leads to (up to unimportant constants)
\begin{align}
{\cal H}=&
	S
	\sum_m
	\biggl(
		J\Bigl(
			\hat{N}_m+\hat{N}_{m+1}
			-
			\hat{N}_m\hat{N}_{m+1}/S
			-
			\hat{a}^\dagger_m\hat{a}_{m+1}
			-
			\hat{a}^\dagger_{m+1}\hat{a}_m
		\Bigr)
		+
		2D
		\Bigl(
			\hat{N}_m
			-
			\hat{N}_m^2/S
		\Bigr)
	\biggr)
\nonumber\\=&
	S
	\sum_m
	\biggl(
		2(J+D)\hat{N}_m
		-
		\frac{1}{S}(D\hat{N}_m+J\hat{N}_{m+1})\hat{N}_m
		-
		J(
			\hat{a}^\dagger_m\hat{a}_{m+1}
			+
			\hat{a}^\dagger_{m+1}\hat{a}_m
		)
	\biggr)
\nonumber\\\approx&
	S
	\sum_m
	\biggl(
		2(J+D)\hat{N}_m
		-
		\frac{1}{S}(J+D)\hat{N}_m^2
		-
		J(
			\hat{a}^\dagger_m\hat{a}_{m+1}
			+
			\hat{a}^\dagger_{m+1}\hat{a}_m
		)
	\biggr)
	,
\end{align}
where the last line is obtained under the assumption that the magnon number $n_m$ varies only slowly with site index. From this expression, there is a direct mapping to Eq. (\ref{BH}) by putting $U=-2(J+D)$, $\mu=-(2S-1)(J+D)$, and $t=SJ$, as well as mapping $m \mapsto p_m$ (prime numbering) with $m>0$.

\section{Summary} We have recast the Bose-Hubbard model in the framework of multiplicative number theory. The specific form of the creation and annihilation operators in this implementation enable error-free computational algorithms. However, it ought to be noted, the algorithms incorporate prime factorization which limits their efficiency. We have subsequently provided rigorous evidence of phase transitions in finite-dimensional implementations of the model. In addition, we have applied the Fourier transform on $\mathbb{Q}_+$ which established the equivalence of number-theoretic structures with the Kastrup model of the harmonic oscillator. This observation spans a conceptual bridge between investigations that have been developed independently and for different purposes, namely \cite{Bost_Connes} on one hand, and \cite{Kastrup} on the other.  The duality also sheds light on the role of the Holstein-Primakoff transform, and bridges the Bose-Hubbard model with a system of spin moments with an anisotropy. We hope the duality highlighted here will inspire more mathematical and physical insights in the future.

\section*{Acknowledgements} The authors are grateful to the anonymous referees for constructive criticism which resulted in improvements to the presentation of these results.


\end{document}